\documentclass[12pt]{article}
\usepackage{epsfig}
\usepackage{amssymb}
\usepackage{cite}

\newcommand\pt{p_{\rm\scriptscriptstyle T}}
\newcommand\hpt{\hat{p}_{\rm\scriptscriptstyle T}}
\newcommand\nlf{{n_{\rm\scriptscriptstyle L}}}
\newcommand\nf{n}
\newcommand\tf{{T_{\rm\scriptscriptstyle F}}}
\newcommand\cf{{C_{\rm\scriptscriptstyle F}}}

\newcommand\as{\alpha_{\rm\scriptscriptstyle S}}
\newcommand\MSB{$\overline{\rm MS}$}

\newcommand\Inlf{\mathbb{I}_\nlf}
\newcommand\Inf{\mathbb{I}_\nf}

\def\beq{\begin{equation}}
\def\eeq{\end{equation}}

\def\beqn{\begin{eqnarray}}
\def\eeqn{\end{eqnarray}}
\def\lq{\left[}
\def\rq{\right]}

\def\({\left(}
\def\){\right)}

\begin{document}
\begin{titlepage}
\nopagebreak
{\flushright{
        \begin{minipage}{4cm}
        Bicocca-FT-05-8\\
        LPTHE-05-09
        \end{minipage}        }

}
\vfill
\begin{center}
{\LARGE 
{ \bf \sc
Crossing Heavy-Flavour Thresholds\\
in Fragmentation Functions 

}}
\vskip .5cm
{\bf Matteo Cacciari}
\\
\vskip .1cm
{LPTHE, Universit\'e Pierre et Marie Curie (Paris 6), France}

\vskip .5cm
{\bf Paolo Nason}
\\
\vskip 0.1cm
{INFN, Sezione di Milano\\
Piazza della Scienza 3, 20126 Milan, Italy} \\

\vskip .5cm
{\bf Carlo Oleari}
\\
\vskip 0.1cm
{Universit\`a di Milano-Bicocca,\\
Piazza della Scienza 3, 20126 Milan, Italy} \\
\end{center}
\nopagebreak
\vfill
\begin{abstract}

In analogy with parton distribution functions, also parton fragmentation
functions obey matching conditions when crossing heavy-flavour thresholds. We
compute these matching conditions at next-to-leading order in the strong
coupling constant $\as$ in the \MSB\ scheme.
Our results can be used for the dynamical generation of the heavy-flavour
component in next-to-leading order fits to light-hadrons fragmentation
functions.  Furthermore, when computing perturbatively the charm
fragmentation function from first principles and evolving it to higher
scales, our matching conditions should be used for consistency when crossing
the bottom threshold.

\end{abstract}
\vskip 1cm
\vfill
\end{titlepage}

\section{Introduction}
In full analogy with structure functions, also fragmentation
functions change when heavy-flavour thresholds are
crossed. In a system with $\nlf$ light flavours, and a heavy
flavour of mass $m$, the inclusive production of a hadron at a scale much
below the heavy-flavour threshold is given by
\begin{equation}\label{eq:factform}
\frac{d\sigma}{d\pt}
=\int dz \, d\hpt \sum_{i\in \Inlf}
 D_i^{(\nlf)}(z,\mu)
\frac{d\hat\sigma_i^{(\nlf)}(\mu)}{d\hpt} \delta(\pt-z\hpt)\;,
\end{equation}
where $\Inlf$ is the set of all light partons
\begin{equation}
\Inlf=\{ q_1,\bar{q}_1,\,\ldots q_\nlf,\bar{q}_\nlf,\,g\}\;,
\end{equation}
${d\hat\sigma_i^{(\nlf)}}/{d\hpt}$ are the short distance cross sections,
calculable in perturbation theory, and $D_i^{(\nlf)}$ are the partonic
fragmentation functions that describe the hadronization of the light partons
$i$ into the hadron.  The partonic fragmentation functions $D_i^{(\nlf)}$ are
normally not calculable perturbatively, but their evolution in terms of the scale
$\mu$ is given by the Altarelli-Parisi equations with $\nlf$ flavours.

Renormalization is performed in the so called decoupling
scheme~\cite{Collins:1978wz}:
divergences associated with light particles are subtracted in the \MSB\ 
scheme, while divergent heavy-quark loop effects are
subtracted at zero momentum. This results in the fact that, for momenta
much below the heavy-quark mass, the heavy-quark existence can
be completely ignored.
In the following, we will also refer to this scheme as the $\nlf$ scheme.

Much above the heavy-flavour threshold, an equation
of the form~(\ref{eq:factform}) also holds, but
with $\nlf$ replaced by $\nf=\nlf+1$.
The $D_i^{(\nf)}$ obey Altarelli-Parisi equations for $\nf$ flavours,
and the standard \MSB\ scheme is used for all flavours.
We will refer to this scheme as the $\nf$ scheme.

For a heavy enough quark, the relation of $D^{(\nf)}$
with $D^{(\nlf)}$ can be computed in perturbative QCD.
At the leading logarithmic level, we have
\begin{equation}\label{eq:matchingll}
D_i^{(\nf)}(z,\mu) =
\left\{ \begin{array}{ll}
 D_i^{(\nlf)}(z,\mu)+{\cal O}(\as)
 &\mbox{for}\; i\in \Inlf \\
{\cal O}(\as)
 &\mbox{for}\; i\in \{h,\bar{h}\} 
\end{array} \right.
\quad \mbox{when}\;\mu\approx m \;,
\end{equation}
where we have introduced the notation $h=q_\nf$ for the heavy flavour.
The matching condition~(\ref{eq:matchingll}) simply states
that the presence of the heavy quark has effects of order $\as$
in processes at scales near its mass.
A relation of the same form as~(\ref{eq:matchingll})
holds for parton densities below and above
a heavy-flavour threshold. In this case, however,
the next-to-leading order (NLO) accurate matching condition has also been
known for a long 
time~\cite{Collins:1986mp}. It  requires
$D^{(\nf)}_h$ to be of order $\as^2$ 
and the difference of $D^{(\nf)}$ and $D^{(\nlf)}$ for
the light flavours to be at most of
order $\as^2$, for $\mu$ exactly equal to $m$.
This matching condition is commonly used in next-to-leading logarithmic (NLL)
parton distribution functions fits, where the heavy-quark distributions
$c(x,\mu)$ and $b(x,\mu)$ are radiatively generated, rather than directly
fitted to the data.

For fragmentation functions, NLO matching has never appeared in the
literature, and it has never been used in global fits.
Usually~\cite{Chiappetta:1992uh,Bourhis:1997yu,Kretzer:2000yf,Bourhis:2000gs,Kniehl:2000fe,Albino:2005me} 
the heavy-flavour fragmentation functions are simply parametrized and fitted
to the data. In~\cite{Coriano:2001rt}, they are instead generated
dynamically, but using only the leading-order matching
condition~(\ref{eq:matchingll}).

In the present work, we compute the matching conditions
for fragmentation functions at the NLO level in the \MSB\ scheme. The NLO
matching conditions should be used for consistency in NLL
fragmentation-function fits that generate heavy flavours dynamically. 
In particular, in the computation of the charm fragmentation
function from first principles~\cite{Mele:1990cw}, the matching
conditions should be used when crossing the bottom threshold.
In this framework, at low energies (i.e.\ not much above the charm mass),
the charm is treated as a heavy quark, in order
to provide a perturbative expression for its fragmentation
function.
Near the bottom threshold, the bottom is treated as heavy,
while all other quarks (including charm) are considered light.

\section{Matching conditions at NLO accuracy}
We consider a light-hadron $H$ inclusive production process
in $e^+e^-$ annihilation\footnote{Universality of the fragmentation
function guarantees that the result obtained in a specific process
remains valid for all processes.}
\beq
e^+e^-\,\to\, H + X
\eeq
at a center-of-mass energy $Q$ much above the
heavy-flavour mass, so that power-suppressed $m/Q$ terms can be neglected,
but not too large,  so that $\alpha_S \log (m/Q) \ll 1$.

In this regime, we can perform the calculation
in both the decoupling and the full \MSB\ scheme.
In the first scheme, we can neglect to resum large logarithms of $m/Q$,
since $\alpha_S \log (m/Q)$ is small, use the $D^{(\nlf)}$
fragmentation functions, and include the heavy-flavour effects
by fixed ${\cal O}(\as)$ calculation. On the other hand, in the second
scheme, we can neglect power-suppressed $m/Q$ terms, treat all flavours
as light, and use the full \MSB\ scheme, with the $D^{(\nf)}$
fragmentation functions and the corresponding factorization formulae.

\begin{figure}[ht]
\begin{center}
\epsfig{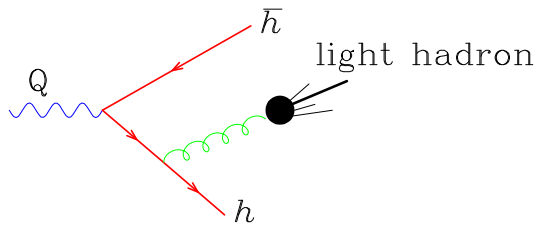}
\end{center}
\caption{\label{heavytog}
Heavy-flavour contribution to the production of a light hadron.}
\label{fig:moments}
\end{figure}

In the decoupling scheme, we can write the differential cross section for the
production of a light hadron $H$ with energy fraction $x=2E_H/Q$ as
\begin{equation} \label{eq:massive}
\frac{d\sigma}{dx}
=\int_x^1 \frac{dy}{y} \left\{
 \sum_{i\in \Inlf} D_i^{(\nlf)}(x/y,\mu)
\frac{d\hat\sigma_i^{(\nlf)}(y,\mu)}{dy}
+ D_g^{(\nlf)}(x/y,\mu)
\frac{d\sigma_{h\bar{h}g}(y)}{dy}\right\} \;.
\end{equation}
The first term involves the fragmentation function
and the short distance cross sections $\hat\sigma_i$ for all partons $i$,
excluding the heavy flavour.
The second term is the ${\cal O}(\as)$ heavy-flavour contribution,
arising from direct production of a heavy-flavoured pair, followed by gluon
radiation and fragmentation, as depicted in Fig.~\ref{heavytog}.
The $\sigma_{h\bar{h}g}$ cross section
can be easily obtained from the appendix of Ref.~\cite{Nason:1994xx}
\begin{equation}\label{eq:sigmahhg}
\frac{d\sigma_{h\bar{h}g}(y)}{dy}=\sigma_{h\bar{h}}\,
\frac{\as}{2\pi}\, \cf \,
2\frac{1+(1-y)^2}{y} \left\{ \log\frac{Q^2}{m^2}+\log(1-y)-1 \right\}\;,
\end{equation}
where $\sigma_{h\bar{h}}$ is the Born cross section for the production
of a heavy quark-antiquark pair,
$y=2E_g/Q$ is the energy fraction carried by the gluon, and $\cf=4/3$.
Terms suppressed by powers of $m/Q$ have been consistently
neglected in Eq.~(\ref{eq:sigmahhg}).

The calculation for the same observable $d\sigma/dx$ can also be performed 
using the full (massless) \MSB\
scheme, treating all flavours as massless. Defining $\Inf =
\Inlf \cup \{h,\bar{h}\}$, we obtain
\begin{equation}\label{eq:msbar}
\frac{d\sigma}{dx}
=\int_x^1 \frac{dy}{y}\sum_{i\in \mathbb{I}_\nf} D_i^{(\nf)}(x/y,\mu)
\frac{d\hat\sigma_i^{(\nf)}(y,\mu)}{dy} \;.
\end{equation}

The difference between the expressions~(\ref{eq:massive})
and~(\ref{eq:msbar}) yields
\begin{eqnarray}
\label{eq:diffs}
&& \!\!\!\!\int_x^1\! \frac{dy}{y}\sum_{i\in \Inlf, \;i\neq g}
\left[D_i^{(\nf)}(x/y,\mu)-D_i^{(\nlf)}(x/y,\mu)\right]
\frac{d\hat\sigma_i(y,\mu)}{dy} 
\nonumber \\ 
&+&\!\!\!\!
 \int_x^1 \!
\frac{dy}{y}\!
\left[D_g^{(\nf)}(x/y,\mu)\frac{d\hat\sigma_g^{(\nf)}(y,\mu)}{dy}   
-D_g^{(\nlf)}(x/y,\mu) \frac{d\hat\sigma_g^{(\nlf)}(y,\mu)}{dy}
\right] 
\nonumber \\ 
&+&\!\!\!\!
\int_x^1 \!\frac{dy}{y} \!\left[
\sum_{i\in \{h,\bar{h}\}} D_i^{(\nf)}(x/y,\mu)
\frac{d\hat\sigma_i(y,\mu)}{dy} -
 D_g^{(\nlf)}(x/y,\mu)
\frac{d\sigma_{h\bar{h}g}(y)}{dy}\right] \!\!=0.
\end{eqnarray}
In the first term of Eq.~(\ref{eq:diffs}) we have replaced
 $\hat\sigma_i^{(\nf)}$ and $\hat\sigma_i^{(\nlf)}$ with $\hat\sigma_i$,
since these cross sections only differ by terms of order $\as^2$.
In the second term, since $\hat{\sigma}_g$ is of order $\as$, we can
consider $D_g^{(\nlf)}=D_g^{(\nf)}$, since their difference,
according to Eq.~(\ref{eq:matchingll}), is of order $\as$.
Furthermore
\begin{equation}
\frac{d\hat\sigma_g^{(\nf)}(y,\mu)}{dy}
-\frac{d\hat\sigma_g^{(\nlf)}(y,\mu)}{dy} =
 \frac{d\hat\sigma_{h\bar{h}g}(y,\mu)}{dy}\;,
\end{equation} 
where $d\hat\sigma_{h\bar{h}g}(y,\mu)/dy$ is the massless \MSB-subtracted cross
section for the process of Fig.~\ref{heavytog}, i.e.\ a gluon emitted from the
(now only nominally) heavy quark.

Finally, in the third term of Eq.~(\ref{eq:diffs}), $D^{(\nf)}_{h/{\bar h}}$
are of order $\as$ (according to Eq.~(\ref{eq:matchingll})), so that we only
need the Born term for the hard cross section
\begin{equation}
\frac{d\hat\sigma_{h/\bar{h}}(y,\mu)}{dy}=\sigma_{h\bar{h}}
\,\delta(1-y)+{\cal O}(\as)\;.
\end{equation}
Since Eq.~(\ref{eq:diffs}) must hold for arbitrary electric charges of each
quark flavour, it follows immediately that, neglecting terms of order
$\as^2$,
\begin{equation}
\label{eq:match_q}
D_i^{(\nf)}(x,\mu)=D_i^{(\nlf)}(x,\mu)\quad\quad\mbox{for}\;
i\in \Inlf, \; i\neq g\;.
\end{equation} 
We thus find
\begin{eqnarray}
&& \int_x^1 \frac{dy}{y}D_g(x/y,\mu)
\left[\frac{d\hat\sigma_{h\bar{h}g}(y,\mu)}{dy} -
\frac{d\sigma_{h\bar{h}g}(y)}{dy}\right]
 \nonumber\\
&& \quad\quad\quad+
\int_x^1 \frac{dy}{y}\,
\sum_{i\in \{h,\bar{h}\}} D_i^{(\nf)}(x/y,\mu)
\frac{d\hat\sigma_i(y,\mu)}{dy} 
\;=\;0\;,
\end{eqnarray}
which implies
\begin{equation}
D_h^{(\nf)}(x,\mu)+D_{\bar{h}}^{(\nf)}(x,\mu)
=\frac{1}{\sigma_{h\bar{h}}} \int_x^1 \frac{dy}{y}\left[
\frac{d\sigma_{h\bar{h}g}(y)}{dy}
-\frac{d\hat\sigma_{h\bar{h}g}(y,\mu)}{dy}\right]\,D_g(x/y,\mu)\;.
\end{equation}
The \MSB\ subtracted cross section $\hat\sigma_{h\bar{h}g}(y,\mu)$ is given
by~\cite{Nason:1994xx, Furmanski:1982cw}
\begin{equation}
\frac{d\hat\sigma_{h\bar{h}g}(y,\mu)}{dy}=\sigma_{h\bar{h}}\,
\frac{\as}{2\pi}\, \cf 2\frac{1+(1-y)^2}{y}
\left\{2 \log y + \log(1-y) + \log\frac{Q^2}{\mu^2} \right\}\;,
\label{eq:hhmsbar}
\end{equation}
and thus, using Eqs.~(\ref{eq:sigmahhg}) and~(\ref{eq:hhmsbar}), we obtain 
\begin{eqnarray}
\label{eq:match_h}
&&D_h^{(\nf)}(x,\mu)= D_{\bar{h}}^{(\nf)}(x,\mu)=\nonumber\\
&& \quad\quad
\int_x^1 \frac{dy}{y}
 \, D_g(x/y,\mu)\, \frac{\as}{2\pi}\, \cf\, 
\frac{1+(1-y)^2}{y}
\left[\log\frac{\mu^2}{m^2}-1-2\log y \right]\!\!.
\end{eqnarray}
Equations~(\ref{eq:match_q}) and~(\ref{eq:match_h}) are the matching
conditions for the $D_q$ and $D_h$, accurate up to terms of order $\as$. 
We remark that the right hand side in Eq.~(\ref{eq:match_h}) does {\em not}
vanish for $\mu = m$. Under this respect, matching conditions for parton 
fragmentation functions differ from those for distribution 
functions~\cite{Collins:1986mp}, 
which instead vanish at the heavy-quark threshold.

\begin{figure}[ht]
\begin{center}
\epsfig{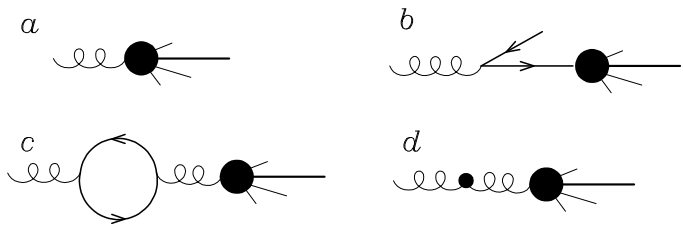}
\end{center}
\caption{\label{gfrag}
Contributions to the production of a light hadron
from a gluon.}
\end{figure}

Contrary to $D_q$ and $D_h$, the matching condition for the gluon
fragmentation function $D_g$ has not yet been determined to NLO accuracy.  In
fact, in $e^+e^-$ annihilation, we are only sensitive to the leading order
value of $D_g$, since gluon production is suppressed by a power of $\as$.
On the other hand, if we consider a wide evolution span from $m$ to $Q$, such
that $\as \log (Q/m) \approx 1$, an ${\cal O}(\as)$ error in $D_g$ at $m$
would propagate into an ${\cal O}(\as)$ error in all the other components of
$D$ through evolution.  We must therefore provide a matching condition for
$D_g$ accurate to order $\as$.  We will show that, in fact,
\begin{equation} \label{eq:gluonmatch}
D_g^{(\nf)}(z,\mu)=D_g^{(\nlf)}(z,\mu) + {\cal O}(\as^2)\quad
\mbox{for}\;\mu=m\;, 
\end{equation}
which is the same result that holds for parton densities~\cite{Collins:1986mp}.
In order to prove Eq.~(\ref{eq:gluonmatch}), we consider
the process of light-hadron production from a gluon.
We can imagine that the gluon is produced in some
physical process. For concreteness, we can think of a ``super-heavy'' quark
produced in $e^+e^-$ annihilation (renormalized in the same way in
the $\nlf$ and $\nf$ scheme), that radiates a gluon.
The gluon produces a light hadron by fragmentation, according
to the graph $a$ in Fig.~\ref{gfrag}.
We now want to consider the difference in the 
radiative corrections to this process in the $\nlf$ and in the $\nf$ scheme.
It is clear that processes where the gluon splits into a pair
of light partons give the same contribution in both schemes.
These are in fact processes of order $\as$, diagrammatically
identical in the $\nlf$ and $\nf$ scheme.
In these processes (that carry an extra power of $\as$)
the strong coupling constant and the fragmentation functions
can be taken at leading order, and therefore they coincide, at ${\cal O}(\as)$,
in the two schemes. The process $g\to h\bar{h}$ followed by the
fragmentation of the heavy quark (shown in graph $b$ of the figure)
exists only in the $\nf$ scheme, but it is suppressed by $\as^2$, since
the fragmentation function $D_h$ is of order  $\as$, and therefore does not
give a relevant contribution. The only corrections that can contribute
to the difference are shown in graphs $c$ and $d$.
In $c$, a heavy quark loop corrects the gluon propagator.
We include in $c$ also the corresponding renormalization counterterm.
Graph $d$ represents the difference in the collinear
subtractions in the $\nf$  and $\nlf$ schemes.

Calling $\Pi$ the contribution from the heavy-quark loop in graph $c$, and
$C_{\rm r}$ the renormalization counterterm, we have
\begin{eqnarray} 
\label{eq:bollauno}
&&\Pi^{(\nlf)}+C_{\rm r}^{(\nlf)} =0\;, \\
\label{eq:bolladue}
&&\Pi^{(\nf)}+ C_{\rm r}^{(\nf)} =
 C^{(\nf)}_{\rm r}=\frac{1}{\epsilon}\frac{\tf \as}{3\pi}\times
\mbox{Born}\;.
\end{eqnarray}
Equation~(\ref{eq:bollauno}) follows from the fact that the
heavy-quark loop is subtracted at zero momentum, and the gluon is on shell
in $c$. Equation~(\ref{eq:bolladue}) follows from the fact
that a massless quark loop at zero momentum is zero in dimensional
regularization, so only the standard counterterm for a quark loop in the
\MSB\ scheme survives. 

For graph $d$, 
the difference in the collinear subtraction counterterm arises
only from the $P_{gg}$ term of the Altarelli-Parisi splitting functions,
which is the only term that depends explicitly upon the number of light
flavours. We have
\begin{equation}
-\frac{1}{\epsilon}
\frac{\as}{2\pi}\left[P^{(\nf)}_{gg}(z)-P^{(\nlf)}_{gg}(z)\right]
=-\frac{1}{\epsilon} \frac{\tf \as}{3\pi}\;,
\end{equation}
that cancels exactly the remnant from Eq.~(\ref{eq:bolladue}).  The
cancellation is in fact obvious. The fermion contribution of the vacuum
polarization from a quark loop vanishes, and thus its ultra-violet (UV)
component must cancel exactly its infra-red (IR) component. The
renormalization counterterm should be equal to the UV component with the
opposite sign, and thus equals the IR component. Finally, the collinear
counterterm arises precisely from the IR component of a quark loop, and so it
compensates the renormalization counterterm.

The only remaining contribution that could differ in the $\nf$ and $\nlf$
schemes is the one represented in graph $a$.
It is proportional to the product of the gluon fragmentation function times
the coupling $\as$ arising from the emission of the gluon.
Since the total result must be scheme independent, we must therefore have
\begin{equation}
\as^{(\nf)}(\mu) D_g^{(\nf)}(z,\mu)=\as^{(\nlf)}(\mu) D_g^{(\nlf)}(z,\mu)\;.
\end{equation}
From the matching condition for the running coupling $\as$
\begin{equation}
\as^{(\nf)} = \as^{(\nlf)} \left(1+\frac{\tf \as}{3\pi}
\log\frac{\mu^2}{m^2}\right)\;, 
\end{equation}
we therefore infer
\begin{equation}
D_g^{(\nf)}(z,\mu)= D_g^{(\nlf)}(z,\mu)
\left(1-\frac{\tf\as}{3\pi} \log\frac{\mu^2}{m^2}\right)\;.
\end{equation}
Summarizing, for $\mu\approx m$ and up to corrections of order
$\as^2$ we have
\begin{eqnarray}
D_{h/\bar{h}}^{(\nf)}(x,\mu) \!\!&=&\!\! \int_x^1 \frac{dy}{y}
 \, D_g(x/y,\mu)\,\nonumber\\
&&\quad\quad\quad\times
\frac{\as}{2\pi}\,\cf\,
\frac{1+(1-y)^2}{y}
\left[\log\frac{\mu^2}{m^2}-1-2\log y\right]\;
\\
 D_g^{(\nf)}(x,\mu)\!\!&=&\!\! D_g^{(\nlf)}(x,\mu)
\left(1-\frac{\tf\as}{3\pi} \log\frac{\mu^2}{m^2}\right)\;
\\
 D_{i/{\bar i}}^{(\nf)}(x,\mu)\!\!&=&\!\! D_{i/{\bar i}}^{(\nlf)}(x,\mu)
\quad\quad\quad\mbox{for}\; i=q_1,\ldots,q_\nlf \;.
\end{eqnarray}
For future reference, we provide the Mellin transform of
$D_{h/\bar{h}}^{(\nf)}$.  Defining 
\beq
D_{h/\bar{h}}^{(\nf)}(N) \equiv \int_0^1 dx \;x^{N-1}
D_{h/\bar{h}}^{(\nf)}(x,\mu) \;,
\eeq
we have
\beqn
\label{eq:D_h_Mellin}
D_{h/\bar{h}}^{(\nf)}(N) &=& \frac{\as}{2\pi}\, \cf  \lq
\frac{2+N+N^2}{N\(N^2-1\)} \( \log\frac{\mu^2}{m^2} -1 \)  
+ \frac{4}{(N-1)^2} -\frac{4}{N^2} \right.
\nonumber\\
&& \quad\quad \quad + \left. \frac{2}{(N+1)^2}\rq  D_g(N)
\eeqn

\section{Conclusions}

In this paper we have calculated, at order $\as$ and in the \MSB\ scheme, 
the matching conditions for parton fragmentation functions at the heavy-quark
thresholds. Such conditions can be used to generate radiatively and to
next-to-leading accuracy charm and
bottom contributions to light-hadron or photon production via fragmentation. 
This is achieved by evolving light-quarks and gluon fragmentation functions 
from a low- to a high-energy scale, and through the heavy-quark thresholds.

In a similar fashion, these matching conditions at next-to-leading accuracy are
required for full consistency when evolving the perturbatively-calculated charm
fragmentation function through the bottom threshold.

The possibility of generating dynamically the heavy-quark contributions to
fragmentation will allow to fit light-hadron fragmentation functions with
less free parameters than presently done, in full analogy with modern parton
distribution functions analyses. We defer these fits to a future paper.

\end{document}